\documentclass[fleqn,12pt,twoside]{article}
\usepackage{espcrc1}

\usepackage{graphicx}
\usepackage[figuresright]{rotating}

\newcommand{\AmS}{{\protect\the\textfont2
  A\kern-.1667em\lower.5ex\hbox{M}\kern-.125emS}}

\usepackage{epsf,wrapfig,color,epsfig}
\def\beq{\begin{equation}}
\def\eeq{\end{equation}}
\def\beqar{\begin{eqnarray}}
\def\eeqar{\end{eqnarray}}

\newcommand\iso[2]{\mbox{${}^{#2}${\rm #1}}}
\def\he#1{\iso{He}{#1}}
\def\be#1{\iso{Be}{#1}}
\def\li#1{\iso{Li}{#1}}
\def\b1#1{\iso{B}{1#1}}

\title{Primordial Nucleosynthesis in the New Cosmology}

\author{R. H. Cyburt\address[UoI]{Department of Physics, University of
Illinois \\
        133 Astronomy Building, 1002 W. Green St MC-221, Urbana, IL
61801 \\ cyburt@uiuc.edu}%
        \thanks{I am grateful for the support provided by the NIC7
organizers and my collaborators B.D Fields and K.A. Olive.  The work
of R.H.C. is supported by the National Science Foundation grant
AST-0092939.}} 

\begin{document}

\maketitle


\begin{abstract}
Big bang nucleosynthesis (BBN) and the cosmic microwave background
(CMB) anisotro- pies independently predict the universal baryon
density.  Comparing their predictions will provide a fundamental test
on cosmology.  Using BBN and the CMB together, we will be able to
constrain particle physics, and predict the primordial, light element
abundances.  These future analyses hinge on new experimental and
observational data.  New experimental data on nuclear cross sections
will help reduce theoretical uncertainties in BBN's predictions.  New
observations of light element abundances will further sharpen BBN's
probe of the baryon density.  Observations from the MAP and PLANCK
satellites will measure the fluctuations in the CMB to unprecedented
accuracy, allowing the precise determination of the baryon density.
When combined, this data will present us with the opportunity to
perform precision cosmology. \footnote[1]{Note in press: with the new
WMAP data, precision cosmology is at hand and the concordance between
BBN and the CMB is tested in astro-ph/0302431} 
\end{abstract}


\section{Introduction}

Big bang nucleosynthesis (BBN) theory predicts the abundances of the
light elements, deuterium (D), helium (\he3 and \he4), and lithium
(\li7) produced during the first three minutes after the big bang.
Anisotropies detected in the cosmic microwave background (CMB) contain
information about the universe at the time of last scattering, about
300,000 years after the big bang.  Both of these pillars of cosmology,
probing two different epochs in the early universe offer independent
determinations of the cosmic baryon density.  Comparing their results
will provide a fundamental test of cosmology.  There is
tentative agreement between current obervations of the CMB
anisotropies and the light element abundances, and their respective
baryon density determinations adding confidence to the cosmic
``standard model'' \cite{cfo1,cfo2,bnt}.

{\begin{minipage}[h]{60mm}
Figure~1. BBN Predictions
\psfig{file=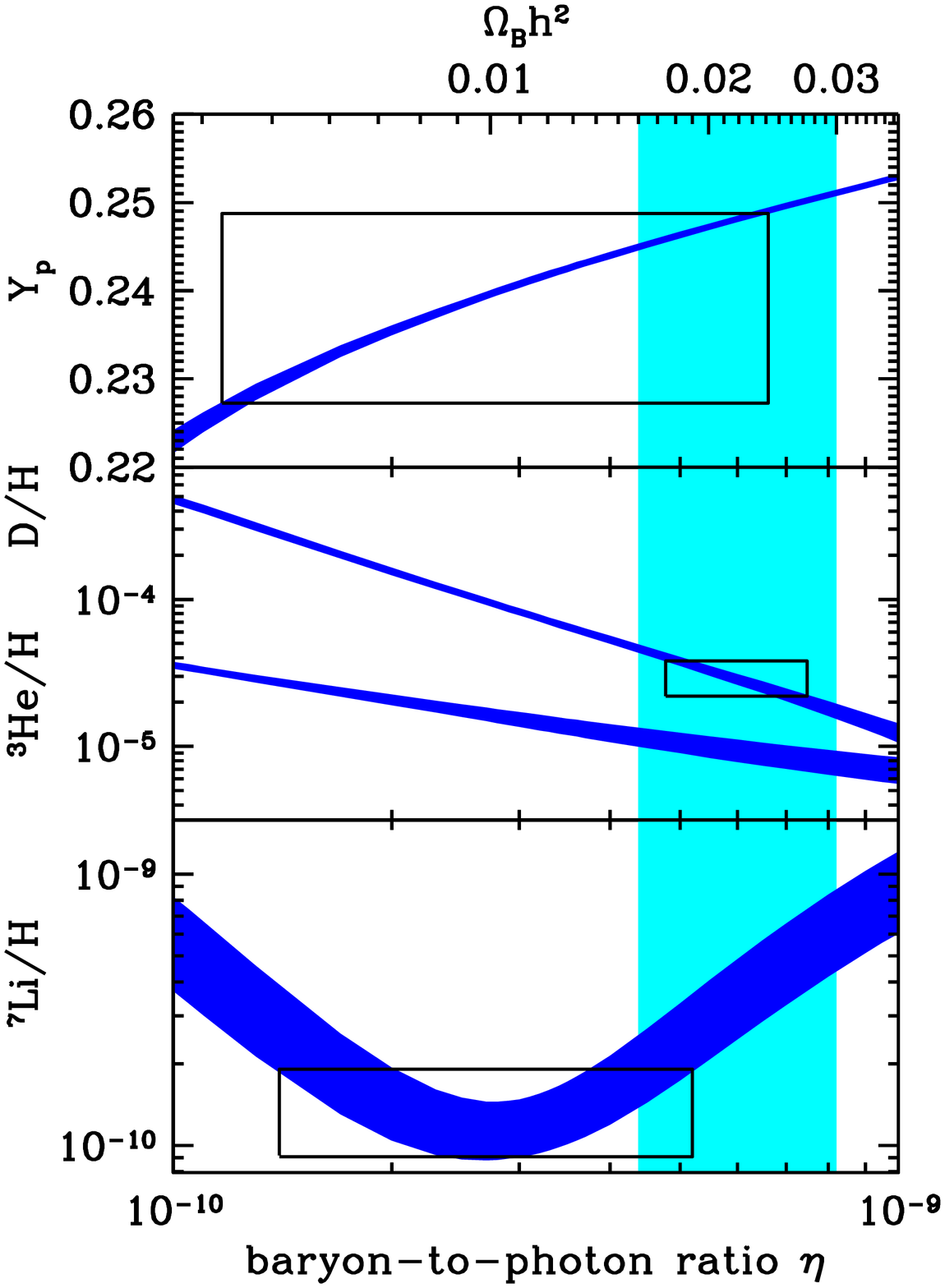,height=3.5in}
\end{minipage}
\hfill
\begin{minipage}[h]{60mm}
\vspace{-10mm}Table~1. Most important reactions and their
uncertainties.
\begin{tabular}[t]{|l||r|} \hline
Reaction  & \% Error \\ \hline\hline
d(p,$\gamma$)\he3                & 13.2 \\ 
\li7(p,$\alpha$)\he4             & 11.4 \\
\he3($\alpha$,$\gamma$)\be7      & 10.6 \\
\he3(d,p)\he4                    & 9.15 \\ \hline\hline
p(n,$\gamma$)d                   & 4.45 \\ 
t($\alpha$,$\gamma$)\li7         & 4.21 \\
t(d,n)\he4                       & 4.01 \\ \hline\hline
\be7(n,p)\li7                    & 3.87 \\ 
\he3(n,p)t                       & 3.52 \\ 
d(d,n)\he3                       & 3.10 \\
d(d,p)t                          & 1.59 \\ 
n$\rightarrow$pe$\bar{\nu}_e$    & 0.09 \\ \hline
\end{tabular}
\end{minipage}}

\section{BBN and the CMB}

 In standard BBN, primordial abundances are
sensitive to only one parameter, the baryon-to-photon ratio, $n_{\rm
B}/n_\gamma \equiv \eta \propto \Omega_{\rm B}h^2$.  Light element
observations are used to determine their primordial abundances
\cite{4obs,Dobs,7obs}.  These in turn, are convolved with BBN theory to
determine the baryon density.  The uncertainties in the theory lie
entirely with the nuclear reaction cross sections that govern the
formation of the light nuclei \cite{cfo1,nuccomp}.  

Figure 1 shows the 95\% confidence predictions
of BBN theory.  Combined with observational constraints, shown with
dark boxes, they pick out an $\eta \sim 5.0\!\times\!\! 10^{-10}$.  Current
CMB measurements by DASI and BOOMERANG \cite{cmb} tentatively agree with
this baryon density, as shown by the vertical band.  

With the
advent of precision measurements of the anisotropies in the CMB by MAP
and PLANCK, we are obliged to go out of our way to improve both theory
and observations in order to make the best comparison between BBN and
the CMB.   

\section{Data Needed}

The reactions listed in Table 1 dominate the
uncertainties in the light element abundance predictions from BBN.
The uncertainties listed are from \cite{cfo1}, see also \cite{nuccomp}
for different and complimentary methods.  It is important to measure
these key reactions to less than 4\% so that BBN theory's predictions are
more precise, making BBN a sharper probe of nuclear and particle
astrophysics \cite{cfo2,sarkar,ichiki}.  

Besides improving the nuclear reaction uncertainties, it is
important to reduce the errors in the light element abundance
observations.  A large number of \he4 and \li7 observations exist from
extragalactic H{\sc ii} regions and the atmospheres of population II
stars in the Galactic halo, respectively.
Both of these abundance determinations suffer from large systematic
uncertainties.  Improvements in theoretical models of these regions
are needed to reduce these
systematics,and possible new obervational stageties should be attempted.

Deuterium however, consists of only a handful of observations.  This
small sample prevents us from analysing systematic uncertainties and
exploring possible trends.  Only with more observations in
high-redshift damped Lyman-$\alpha$ systems will D be put on the same
statistical footing as \he4 and \li7.

\section{BBN with the CMB}

With the direct
comparision of the baryon densities derived from BBN and the CMB, we
will fundamentally test the very framework cosmology is built upon.
Assuming concordance is established we can use BBN and
the CMB together to place stronger constraints on physics
\cite{cfo1,cfo2}.  With the CMB baryon density, we can predict the
light element abundances with high precision.  Using the observations
of light element abundances we can address their evolution in
universe.  We can test new particle physics (reviewed in
\cite{sarkar}) and non-standard cosmology (e.g. dark radiation \cite{ichiki}).
With new analyses of observational data determining the primordial
light element abundances and updated nuclear cross section data, we
will be able to perform precision cosmology.  

\section{Acknowledgments}

I would like thank Timothy Beers, Sean Ryan, Jason Prochaska, John Norris,
and Scott Burles for useful discussions about the light element
observations.



\end{document}